\documentclass[a4paper]{report}
\usepackage[utf8]{inputenc}
\usepackage[T1]{fontenc}
\usepackage{RJournal}
\usepackage{amsmath,amssymb,array}
\usepackage{booktabs}

\usepackage{longtable}

\makeatletter
 \let\@cite@ofmt\@firstofone
 \def\@biblabel#1{}
 \def\@cite#1#2{{#1\if@tempswa , #2\fi}}
\makeatother
\newlength{\cslhangindent}
\newlength{\csllabelwidth}
 {\begin{list}{}{%
  \setlength{\itemindent}{0pt}
  \setlength{\leftmargin}{0pt}
  \setlength{\parsep}{0pt}
  \ifodd #1
   \setlength{\leftmargin}{\cslhangindent}
   \setlength{\itemindent}{-1\cslhangindent}
  \fi
  \setlength{\itemsep}{#2\baselineskip}}}
 {\end{list}}
\usepackage{calc}

\usepackage{amsmath} \usepackage{array} \usepackage{float}

\begin{document}

\sectionhead{Contributed research article}
\volume{XX}
\volnumber{YY}
\year{20ZZ}
\month{AAAA}

\begin{article}
\title{quollr: An R Package for Visualizing 2-D Models from Nonlinear Dimension Reductions in High-Dimensional Space}

\author{by Jayani P. Gamage, Dianne Cook, Paul Harrison, Michael Lydeamore, and Thiyanga S. Talagala}

\maketitle

\abstract{%
Nonlinear dimension reduction methods provide a low-dimensional representation of high-dimensional data by applying a Nonlinear transformation. However, the complexity of the transformations and data structures can create wildly different representations depending on the method and hyper-parameter choices. It is difficult to determine whether any of these representations are accurate, which one is the best, or whether they have missed important structures. The R package quollr has been developed as a new visual tool to determine which method and which hyper-parameter choices provide the most accurate representation of high-dimensional data. The scurve data from the package is used to illustrate the algorithm. Single-cell RNA sequencing (scRNA-seq) data from mouse limb muscles are used to demonstrate the usability of the package.
}

\section{Introduction}\label{introduction}

Nonlinear dimension reduction (NLDR) techniques, such as t-distributed stochastic neighbor embedding (tSNE) \citep{laurens2008}, uniform manifold approximation and projection (UMAP) \citep{leland2018}, potential of heat-diffusion for affinity-based trajectory embedding (PHATE) algorithm \citep{moon2019}, large-scale dimensionality reduction Using triplets (TriMAP) \citep{amid2019}, and pairwise controlled manifold approximation (PaCMAP) \citep{yingfan2021}, can create hugely different representations depending on the selected method and hyper-parameter choices. It is difficult to determine whether any of these representations are accurate, which one is the best, or whether they have missed important structures.

\begin{figure}[!ht]
\includegraphics[width=1\linewidth,alt={The UMAP layout with a triangular mesh on the left and the tour view of the lifted wireframe overlaying the data on the right. The mesh illustrates how local neighborhoods in the 2-D embedding correspond to curved sheet structures in the original high-dimensional S-curve data.}]{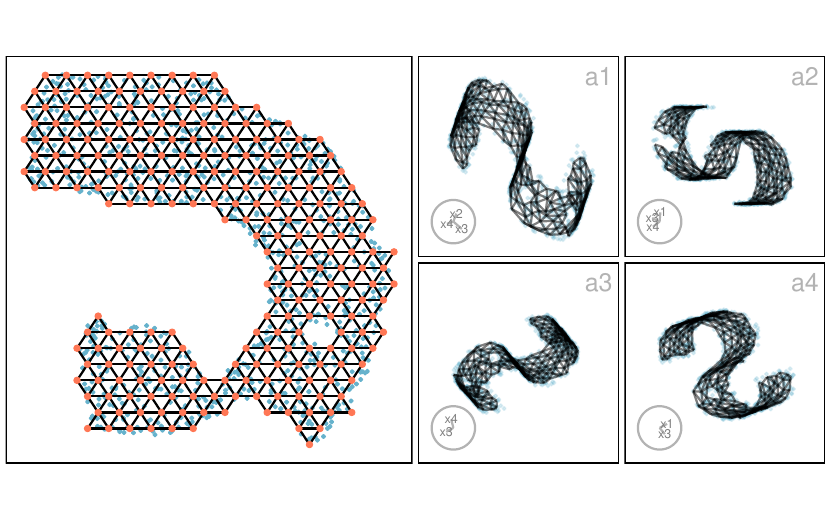} \caption{Wireframe model representation of the NLDR layout, lifted and displayed in high-dimensional space. The left panel shows the NLDR layout with a triangular mesh overlay, forming the wireframe structure. This mesh can be lifted into higher dimensions and projected to examine how the geometric structure of the data is preserved. Panels (a1–a4) display different $2\text{-}D$ projections of the lifted wireframe, where the underlying curved sheet structure of the data is more clearly visible. The triangulated mesh highlights how local neighborhoods in the layout correspond to relationships in the high-dimensional space, enabling diagnostics of distortion and preservation across dimensions.}\label{fig:overview}
\end{figure}

This paper presents the R package, \texttt{quollr}, which is useful for understanding how NLDR warps high-dimensional space and fits the data. Starting with an NLDR layout, our approach is to create a \(2\text{-}D\) wireframe model representation, that can be lifted and displayed in the high-dimensional (\(p\text{-}D\)) space (Figure \ref{fig:overview}).

The paper is organized as follows. The usage section explains how users can fit a complete model pipeline from hexagonal binning of a \(2\text{-}D\) layout to lifting the representation back into \(p\text{-}D\) and how the resulting objects can be explored interactively using tour. The next section introduces the implementation of the \texttt{quollr} package on CRAN, including a demonstration of the package's key functions and visualization capabilities. In the application section, we illustrate the algorithm's functionality for studying a clustering data structure. Finally, we conclude the paper with a brief summary and discuss potential opportunities for using our algorithm.

\section{Usage}\label{usage}

The package is available on CRAN, and the development version is available at \href{https://github.com/JayaniLakshika/quollr}{github.com/JayaniLakshika/quollr}.

Our algorithm includes the following steps: (1) scaling the NLDR data, (2) computing configurations of a hexagon grid, (3) binning the data, (4) obtaining the centroids of each bin, (5) indicating neighboring bins with line segments that connect the centroids, and (6) lifting the model into high dimensions. A detailed description of the algorithm can be found in \citet{gamage2025c}.

The user needs two inputs: the high-dimensional dataset and the corresponding NLDR data. The high-dimensional data must contain a unique \texttt{ID} column, with data columns prefixed by the letter \texttt{x} (e.g., \texttt{x1}, \texttt{x2}, etc.). The NLDR dataset should include embedding coordinates labeled as \texttt{emb1} and \texttt{emb2}, ensuring one-to-one correspondence with the high-dimensional data through the shared \texttt{ID}. The built-in example datasets, \texttt{scurve} and \texttt{scurve\_umap} demonstrates these structures.

To run the entire model pipeline, we can use the \texttt{fit\_high\_model()} function. This function requires: the high-dimensional data (\texttt{highd\_data}), the embedding data (\texttt{nldr\_data}), the number of bins along the x-axis (\texttt{b1}), the buffer amount as a proportion of the data (\texttt{q}), and a benchmark value to identify high-density hexagons (\texttt{hd\_thresh}).

The function returns an object of class \texttt{highd\_vis\_model} containing the scaled NLDR object (\texttt{nldr\_scaled\_obj}) with three elements: the first is the scaled NLDR data (\texttt{scaled\_nldr}), and the second and third are the limits of the original NLDR data (\texttt{lim1} and \texttt{lim2}); the hexagonal object (\texttt{hb\_obj}), the fitted model in both \(2\text{-}D\) (\texttt{model\_2d}), and \(p\text{-}D\) (\texttt{model\_highd}), and triangular mesh (\texttt{trimesh\_data}).

\begin{verbatim}
model_obj <- fit_highd_model(
  highd_data = scurve,
  nldr_data = scurve_umap,
  b1 = 21,
  q = 0.1,
  hd_thresh = 0)
\end{verbatim}

\begin{figure}[!ht]

{\centering \includegraphics[width=0.8\linewidth,alt={The figure is a multi-panel diagram illustrating four steps in constructing a model on a UMAP layout using the scurve data. Panel (a) shows a 2-D UMAP scatter plot with points grouped into hexagonal bins covering the layout. Panel (b) shows the same layout with a centroid point displayed for each hexagon bin. Panel (c) shows the centroids connected by straight line segments to form a triangulated mesh across the 2-D layout. Panel (d) shows this 2-D triangulated mesh lifted into the original high-dimensional space, where the mesh appears as a curved surface. The horizontal and vertical axes in the 2-D panels represent the two UMAP coordinates with bounded numeric ranges, while the lifted view represents coordinates in the higher-dimensional space without explicit axis scales.}]{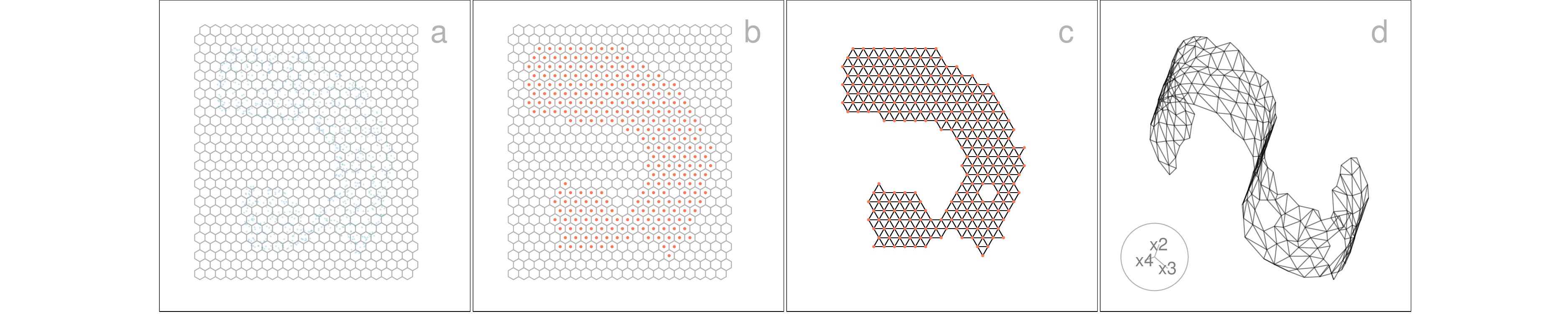}

}

\caption{Key steps for constructing the model on the UMAP layout: (a) hexagon bins, (b) bin centroids, (c) triangulated centroids, and (d) lifting the model into high dimensions. The \texttt{scurve} data is shown.}\label{fig:algo-steps}
\end{figure}

The resulting model can then be shown in a tour using a two-step process:

\begin{verbatim}
combined_data <- comb_data_model(
  highd_data = scurve,
  model_highd = model_obj$model_highd,
  model_2d = model_obj$model_2d
)

tour_view <- show_langevitour(
  point_data = combined_data,
  edge_data = model_obj$trimesh_data
)
\end{verbatim}

which produces the model and data plot shown in Figure \ref{fig:algo-steps}.

\section{Implementation}\label{implementation}

\begin{figure}[H]
\includegraphics[width=1\linewidth,alt={The figure is a schematic diagram illustrating the quollr workflow and software architecture. Rectangular boxes represent processing steps and data objects, including NLDR and p-D data inputs, data preprocessing, hexagonal binning, centroid computation, triangulation, and construction of a 2-D mesh. Arrows indicate the flow of information between steps, from the input data through the 2-D mesh, and then from the mesh being lifted into the p-D space. Additional boxes show prediction and error computation for new data, as well as interactive components that link the p-D and 2-D representations. The diagram is laid out sequentially, showing how data move through the system from input to interactive output.}]{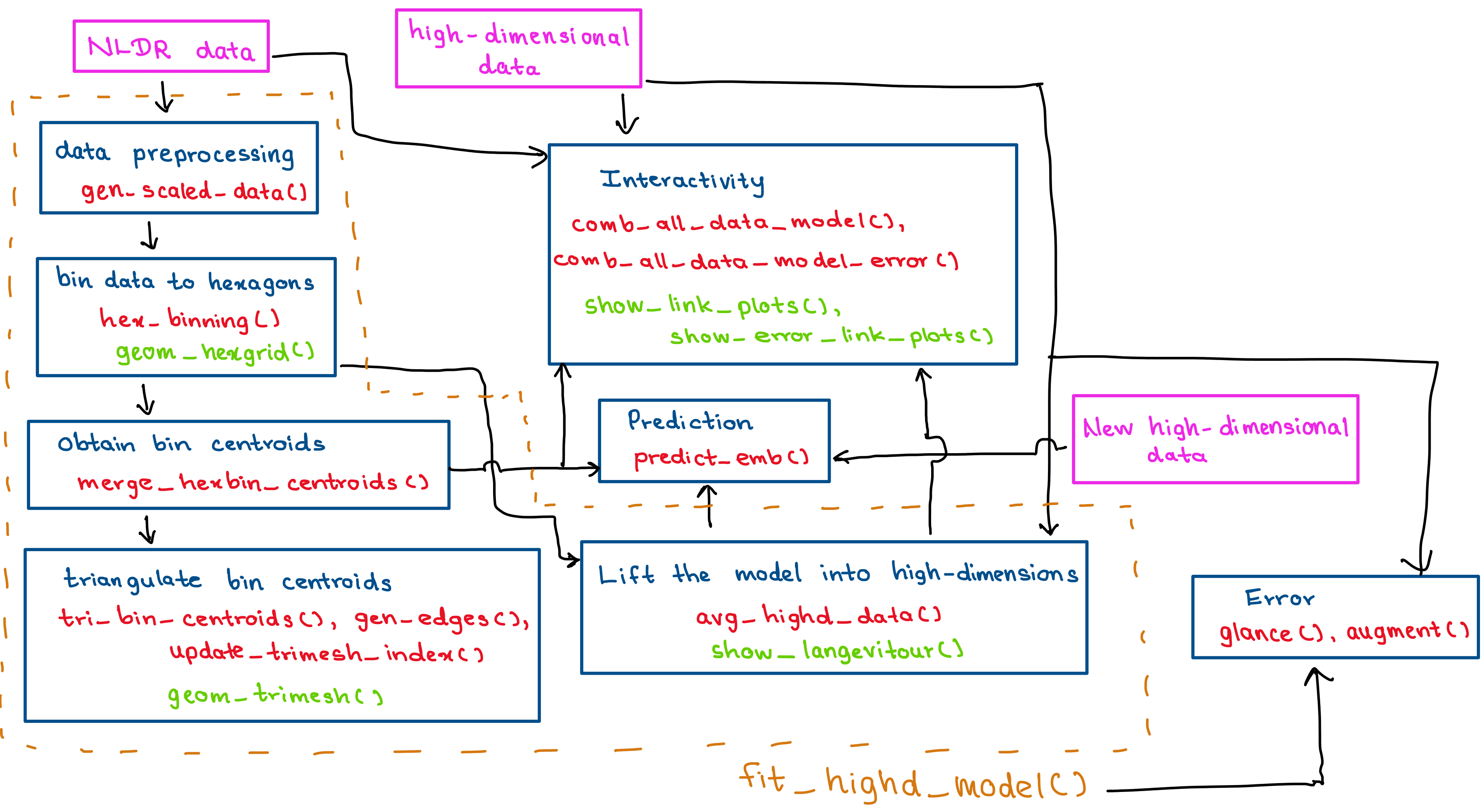} \caption{Overview of the \texttt{quollr} workflow and software architecture. The process begins with NLDR and $p\text{-}D$ data inputs, followed by data preprocessing and hexagonal binning. Centroids are computed and triangulated to form the $2\text{-}D$ mesh, which is then lifted into the $p\text{-}D$ space. Predictions and error computations are performed on new data, while interactive functions enable dynamic linking between the $p\text{-}D$ and $2\text{-}D$ representations.}\label{fig:workflow}
\end{figure}

The implementation of \texttt{quollr} is designed to be efficient, and easy to extend. The package is organized into a series of logical components that reflect the main stages of the workflow: data preprocessing, model fitting, low-density bin removal, prediction, visualization, and interactive exploration (Figure \ref{fig:workflow}). This package structure makes the code easier to maintain and allows new features to be added without changing the existing functionality.

\subsection{Software architecture}\label{software-architecture}

The package is organized into seven core modules corresponding to stages of the workflow: preprocessing, \(2\text{-}D\) model construction, lifting the model into \(p\text{-}D\), prediction, error computation, visualizations, and interactivity. Each module performs a distinct task and communicates through data objects.

\begin{enumerate}
\def\labelenumi{\arabic{enumi}.}
\item
  Data preprocessing: The function \texttt{gen\_scaled\_data()} standardizes the embedding data, manage variable naming, and ensure consistent identifiers across high-dimensional and embedded datasets.
\item
  Construct \(2\text{-}D\) model: A series of functions \texttt{hex\_binning()}, \texttt{merge\_hexbin\_centroids()}, \texttt{tri\_bin\_centroids()}, \texttt{gen\_edges()}, and \texttt{update\_trimesh\_index()} generate the hexagonal grid, compute bin centroids, and connect the triangular mesh that defines local neighborhoods in the \(2\text{-}D\) space.
\item
  Lift the model into \(p\text{-}D\): The function \texttt{avg\_highd\_data()} computes the average of the high-dimensional variables for each bin, linking the \(2\text{-}D\) representation back to the original data space.
\item
  Prediction: The function \texttt{predict\_emb()} estimates the embedding of new high-dimensional observations based on the fitted model.
\item
  Error computation: The \texttt{glance()} and \texttt{augment()} function summarizes model performance by comparing the predicted and original embeddings.
\item
  Visualizations: Functions such as \texttt{geom\_hexgrid()}, \texttt{geom\_trimesh()}, and \texttt{show\_langevitour()} provide tools for exploring the fitted models through static and dynamic visualizations.
\item
  Interactivity: The functions \texttt{comb\_all\_data\_model()} and \texttt{show\_link\_plots()} generate interactive linked visualizations that connect the \(2\text{-}D\) NLDR layout, the corresponding tour view, and the fitted model. Similarly, \texttt{comb\_all\_data\_model\_error()} and \texttt{show\_error\_link\_plots()} integrate the error distribution with the \(2\text{-}D\) embedding and tour view, enabling interactivity across multiple plots.
\end{enumerate}

Each module is internally independent but connected through data objects (see next section). This modular design simplifies maintenance and allows developers to extend individual components such as substituting different binning approaches, extracting centroids, or visualization tools without altering the overall workflow.

\subsection{Data objects}\label{data-objects}

The internal data objects follow the tidy data principle: each variable is stored in a column, each observation in a row, and each type of information in its own table. This structure makes the package easy to use with the \texttt{tidyverse} and other visualization tools.

\subsubsection{Input objects}\label{input-objects}

\begin{itemize}
\item
  \texttt{highd\_data}: a tibble containing the original high-dimensional observations with a unique identifier (\texttt{ID}) and variable columns prefixed with \texttt{x} (e.g., \texttt{x1}, \texttt{x2}, \ldots).
\item
  \texttt{nldr\_data}: a tibble containing two-dimensional embeddings, labeled as \texttt{emb1} and \texttt{emb2}, matched to the same \texttt{ID}s.
\end{itemize}

\subsubsection{Generated objects}\label{generated-objects}

\begin{itemize}
\item
  \texttt{scaled\_nldr\_obj}: the output of \texttt{gen\_scaled\_data()}, which rescales the embedding to the range \([0, 1] \times [0, y_{2,\text{max}}]\), where \(y_{2,\text{max}} = r_2 / r_1\) is the ratio of the embedding ranges. It includes the scaled coordinates (\texttt{scaled\_nldr}) and the original limits (\texttt{lim1}, \texttt{lim2}).
\item
  \texttt{hex\_bin\_obj}: the object created by \texttt{hex\_binning()}, which includes hexagon grid configurations. It includes the binwidth (\texttt{a1}), binheight (\texttt{a2}), the number of bins along each axis (\texttt{b1}, \texttt{b2}), the centroids of all hexagons, polygon coordinates, and the assignment of each data point to the hexagon.
\item
  \texttt{highd\_vis\_model}: the main model object returned by \texttt{fit\_highd\_model()}. It stores all components of the fitted model, including the scaled NLDR data (\texttt{nldr\_scaled\_obj}), the hexagonal bin configurations (\texttt{hb\_obj}), the averaged \(p\text{-}D\) summaries for each bin (\texttt{model\_highd}), the corresponding \(2\text{-}D\) bin centroids (\texttt{model\_2d}), and the triangulated mesh connecting neighboring bins (\texttt{trimesh\_data}).
\end{itemize}

\subsection{Computational efficiency and optimization}\label{computational-efficiency-and-optimization}

Several core computations within \texttt{quollr} are optimized using compiled C++ code via the \texttt{Rcpp} and \texttt{RcppArmadillo} packages. While the user interacts with high-level R functions, performance-critical steps such as nearest-neighbor searches (\texttt{compute\_highd\_dist()}), error metrics (\texttt{compute\_errors()}), \(2\text{-}D\) distance calculations (\texttt{calc\_2d\_dist\_cpp()}), and generation of hexagon coordinates (\texttt{gen\_hex\_coord\_cpp()}) are handled internally in C++. This design provides significant speedups when analyzing large datasets while maintaining a user-friendly R interface. These C++ functions are not exported but are bundled within the package and fully accessible for inspection in the source code.

\subsection{Pipeline implementation}\label{pipeline-implementation}

In this section, we demonstrate the implementation of each step of the pipeline discussed in the Usage section (\texttt{fit\_highd\_model} and \texttt{comb\_data\_model}). Each step can be run independently to ensure flexibility in the modelling aproach.

The algorithm starts by scaling the NLDR data to to the range \([0, 1] \times [0, y_{2,max}]\), where \(y_{2,max} = r_2/r_1\) is the ratio of ranges of embedding components. The output includes the scaled NLDR data (\texttt{scaled\_nldr}) along with the original limits of the embeddings (\texttt{lim1}, \texttt{lim2}).

\begin{verbatim}
scurve_umap_obj <- gen_scaled_data(nldr_data = scurve_umap)
\end{verbatim}

\subsubsection{Hexagon binning}\label{hexagon-binning}

The hexagon binning process builds a complete hexagonal tessellation over the \(2\text{-}D\) NLDR embedding and assigns points to bins for downstream modeling. The workflow proceeds through several steps, beginning with selecting an appropriate grid configurations and ending with assigning NLDR points to their corresponding hexagons.

The first step is to determine the configuration of the hexagonal grid. The function \texttt{calc\_bins\_y()} computes the number of bins along the y-axis (\texttt{b2}), the hexagon width (\texttt{a1}), and the hexagon height (\texttt{a2}), based on three inputs: (i) the scaled NLDR object (\texttt{nldr\_scaled\_obj}, containing the scaled embedding and its \(2\text{-}D\) limits), (ii) the chosen number of bins along the x-axis (\texttt{b1}), and (iii) a buffer proportion (\texttt{q}) that expands the grid slightly beyond the observed data range. This buffer ensures that the tessellation fully covers the embedding. By default, \texttt{q\ =\ 0.1}, but in practice it must be chosen to avoid exceeding the domain of the scaled NLDR data.

\begin{verbatim}
bin_configs <- calc_bins_y(
  nldr_scaled_obj = scurve_umap_obj,
  b1 = 21,
  q = 0.1)

bin_configs
\end{verbatim}

\begin{verbatim}
> $b2
> [1] 28
>
> $a1
> [1] 0.05869649
>
> $a2
> [1] 0.05083265
\end{verbatim}

Once the grid configurations are known, the next step is to construct the full set of hexagon centroids. The function \texttt{gen\_centroids()} uses the bin configuration and the embedding limits to determine the horizontal and vertical spacing of rows, including the staggered row offsets required for hexagonal tiling. Odd rows place centroids directly above one another, while even rows shift by half the hexagon width to create the interlocking structure. Vertical spacing is determined by \(v_s = (\sqrt{3}/2) a_1\), with starting positions defined by the buffer-adjusted limits. This produces a complete grid of potential hexagons covering the \(2\text{-}D\) space.

\begin{verbatim}
all_centroids_df <- gen_centroids(
  nldr_scaled_obj = scurve_umap_obj,
  b1 = 21,
  q = 0.1
)

head(all_centroids_df, 5)
\end{verbatim}

\begin{verbatim}
> # A tibble: 5 x 3
>       h     c_x    c_y
>   <int>   <dbl>  <dbl>
> 1     1 -0.1    -0.116
> 2     2 -0.0413 -0.116
> 3     3  0.0174 -0.116
> 4     4  0.0761 -0.116
> 5     5  0.135  -0.116
\end{verbatim}

After determining the centroid locations, each hexagon's polygonal boundary is generated using the \texttt{gen\_hex\_coord()} function, which constructs the six vertices of every hexagon by applying fixed geometric offsets derived from the hexagon width, \(a_1\). Specifically, the horizontal and vertical offsets are defined as \(dx = a_1/2\), \(dy = a_1/\sqrt{3}\), and \(vf = a_1/(2\sqrt{3})\), and these constants are used to position each vertex relative to the centroid, producing the complete set of hexagon boundaries.

These constants define how far each vertex lies from the centroid in the six characteristic directions. For efficiency, the vertex generation is implemented in C++ and returns a tibble listing all polygon vertices, uniquely indexed by hexagons.

\begin{verbatim}
all_hex_coord <- gen_hex_coord(
  centroids_data = all_centroids_df,
  a1 = bin_configs$a1
)

head(all_hex_coord, 5)
\end{verbatim}

\begin{verbatim}
>   h           x           y
> 1 1 -0.10000000 -0.08179171
> 2 1 -0.12934824 -0.09873593
> 3 1 -0.12934824 -0.13262436
> 4 1 -0.10000000 -0.14956858
> 5 1 -0.07065176 -0.13262436
\end{verbatim}

The next step is to assign each NLDR point to its closest hexagon. The \texttt{assign\_data()} function computes all pairwise Euclidean distances between the \(2\text{-}D\) NLDR embedding and the centroid coordinates, identifying the nearest centroid for every observation. Each point is assigned a hexagon index (\texttt{h}), producing a version of the embedding annotated with bin membership.

\begin{verbatim}
umap_hex_id <- assign_data(
  nldr_scaled_obj = scurve_umap_obj,
  centroids_data = all_centroids_df
)

head(umap_hex_id, 5)
\end{verbatim}

\begin{verbatim}
> # A tibble: 5 x 4
>    emb1  emb2    ID     h
>   <dbl> <dbl> <int> <int>
> 1 0.277 0.913     1   427
> 2 0.697 0.538     2   287
> 3 0.779 0.399     3   226
> 4 0.173 0.953     4   446
> 5 0.218 0.983     5   468
\end{verbatim}

Finally, the list of points within each hexagon can be extracted using \texttt{group\_hex\_pts()}. This step collapses the point-level assignments into a hexagon-level summary by grouping by \texttt{h} and collecting the IDs of all points belonging to that polygon. The result is a tidy representation of the binning structure.

\begin{verbatim}
pts_df <- group_hex_pts(
  scaled_nldr_hexid = umap_hex_id
)

head(pts_df, 5)
\end{verbatim}

\begin{verbatim}
> # A tibble: 5 x 2
>       h pts_list
>   <int> <list>
> 1    58 <int [4]>
> 2    68 <int [1]>
> 3    69 <int [5]>
> 4    70 <int [6]>
> 5    71 <int [9]>
\end{verbatim}

Although each component of the workflow can be run independently, the \texttt{hex\_binning()} function automates the entire sequence---from grid construction to point assignment---and returns a \texttt{hex\_bin\_obj} containing the bin dimensions (\texttt{a1}, \texttt{a2}), the full grid geometry (\texttt{centroids}, \texttt{hex\_poly}), bin membership (\texttt{data\_hb\_id}), standardized counts (\texttt{std\_cts}), the total number of bins (\texttt{b}), non-empty bins (\texttt{m}), and the list of points per bin (\texttt{pts\_bins}).

\begin{verbatim}
hb_obj <- hex_binning(
  nldr_scaled_obj = scurve_umap_obj,
  b1 = 21,
  q = 0.1)
\end{verbatim}

\subsubsection{Computing the standardized number of points within each hexagon}\label{computing-the-standardized-number-of-points-within-each-hexagon}

The \texttt{compute\_std\_counts()} function calculates both the raw and standardized counts of points inside each hexagon.

The function begins by grouping the data by hexagon (\texttt{h}) and counting the number of NLDR points falling within each bin. These raw counts are stored as \texttt{n\_h}. To enable comparisons across bins with varying densities, the function then standardizes these counts by dividing each bin's count by the maximum count across all bins. This yields a standardized bin counts, \texttt{w\_h}, ranging from \(0\) to \(1\).

\begin{verbatim}
std_df <- compute_std_counts(
  scaled_nldr_h = umap_hex_id
  )

head(std_df, 5)
\end{verbatim}

\begin{verbatim}
> # A tibble: 5 x 3
>       h   n_h   w_h
>   <int> <int> <dbl>
> 1    58     4 0.004
> 2    68     1 0.001
> 3    69     5 0.005
> 4    70     6 0.006
> 5    71     9 0.009
\end{verbatim}

\subsubsection{Obtaining bin centroids}\label{obtaining-bin-centroids}

The \texttt{merge\_hexbin\_centroids()} function combines hexagonal bin coordinates, raw and standardized counts within each hexagons.

This function performs a full join with \texttt{centroids\_data}, aligning hexagons (\texttt{h}) between the two datasets to incorporate both hexagonal bin centroids (\texttt{h}) and count metrics. After merging, the function handles missing values in the count columns: any \texttt{NA} values in \texttt{w\_h} or \texttt{n\_h} are replaced with zeros. This ensures that hexagons with no assigned data points are retained in the output, with zero values for count-related fields. The resulting data contains the full set of hexagon centroids along with associated bin counts (\texttt{n\_h}) and standardized counts (\texttt{w\_h}).

\begin{verbatim}
df_bin_centroids <- merge_hexbin_centroids(
  centroids_data = all_centroids_df,
  counts_data = hb_obj$std_cts
  )

head(df_bin_centroids, 5)
\end{verbatim}

\begin{verbatim}
>   h         c_x        c_y n_h w_h
> 1 1 -0.10000000 -0.1156801   0   0
> 2 2 -0.04130351 -0.1156801   0   0
> 3 3  0.01739298 -0.1156801   0   0
> 4 4  0.07608947 -0.1156801   0   0
> 5 5  0.13478596 -0.1156801   0   0
\end{verbatim}

\subsubsection{Indicating neighbors by line segments connecting centroids}\label{indicating-neighbors-by-line-segments-connecting-centroids}

To represent the neighborhood structure of hexagonal bins in a \(2\text{-}D\) layout, we employ Delaunay triangulation \citep{lee1980, albrecht2024} on the centroids of hexagons.

The \texttt{tri\_bin\_centroids()} function generates a triangulation object from the x and y coordinates of hexagon centroids using the \texttt{interp::tri.mesh()} function \citep{albrecht2024}.

\begin{verbatim}
tr_object <- tri_bin_centroids(
  centroids_data = df_bin_centroids
  )
\end{verbatim}

The \texttt{gen\_edges()} function uses this triangulation object to extract line segments between neighboring bins. It constructs a unique set of bin-to-bin connections by identifying the triangle edges and filtering duplicate or reversed links. Each edge is then annotated with its start and end coordinates, and a Euclidean distance between the coordinates.

\begin{verbatim}
trimesh <- gen_edges(tri_object = tr_object, a1 = hb_obj$a1)

head(trimesh, 5)
\end{verbatim}

\begin{verbatim}
> # A tibble: 5 x 8
>    from    to  x_from  y_from    x_to    y_to from_count to_count
>   <int> <int>   <dbl>   <dbl>   <dbl>   <dbl>      <dbl>    <dbl>
> 1     1     2 -0.1    -0.116  -0.0413 -0.116           0        0
> 2    22    23 -0.0707 -0.0648 -0.0120 -0.0648          0        0
> 3    22    44 -0.0707 -0.0648 -0.0413 -0.0140          0        0
> 4     3    23  0.0174 -0.116  -0.0120 -0.0648          0        0
> 5    44    45 -0.0413 -0.0140  0.0174 -0.0140          0        0
\end{verbatim}

The \texttt{update\_trimesh\_index()} function re-indexes the node IDs to ensure that edge identifiers are sequentially numbered and consistent with downstream analysis. This sequential ordering of edges is essential because software such as \texttt{langevitour} and \texttt{detourr} rely on one-to-one mapping between edges and their corresponding vertices to visualize the mesh.

\begin{verbatim}
trimesh <- update_trimesh_index(trimesh_data = trimesh)

head(trimesh, 5)
\end{verbatim}

\begin{verbatim}
> # A tibble: 5 x 10
>    from    to  x_from  y_from    x_to    y_to from_count to_count from_reindexed
>   <int> <int>   <dbl>   <dbl>   <dbl>   <dbl>      <dbl>    <dbl>          <int>
> 1     1     2 -0.1    -0.116  -0.0413 -0.116           0        0              1
> 2    22    23 -0.0707 -0.0648 -0.0120 -0.0648          0        0             22
> 3    22    44 -0.0707 -0.0648 -0.0413 -0.0140          0        0             22
> 4     3    23  0.0174 -0.116  -0.0120 -0.0648          0        0              3
> 5    44    45 -0.0413 -0.0140  0.0174 -0.0140          0        0             44
> # i 1 more variable: to_reindexed <int>
\end{verbatim}

\subsubsection{Identifying and removing low-density hexagons}\label{identifying-and-removing-low-density-hexagons}

Not all hexagons contain meaningful information. Some may have very few or no data points due to the sparsity or shape of the underlying structure. Simply removing hexagons with low counts (e.g., fewer than a fixed threshold) can lead to gaps or ``holes'' in the \(2\text{-}D\) structure, potentially disrupting the continuity of the representation.

To address this, we propose a more nuanced method that evaluates each hexagon not only based on its own density, but also in the context of its immediate neighbors. The \texttt{find\_low\_dens\_hex()} function identifies hexagonal bins with insufficient local support by calculating the average standardized count across their six neighboring bins. If this mean neighborhood density is below a user-defined threshold (e.g., \(0.05\)), the hexagon is flagged for removal.

The \texttt{find\_low\_dens\_hex()} function relies on a helper, \texttt{compute\_mean\_density\_hex()}, which iterates over all hexagons and computes the average density across neighbors based on their hexagon (\texttt{h}) and a defined number of bins along the x-axis (\texttt{b1}). The hexagonal layout assumes a fixed grid structure, so neighbor IDs are computed by positional offsets.

\begin{verbatim}
low_density_hex <- find_low_dens_hex(
  model_2d = df_bin_centroids,
  b1 = 21,
  md_thresh = 0.05
)
\end{verbatim}

For simplicity, we remove low-density hexagons using a threshold of \(0\).

\begin{verbatim}
df_bin_centroids <- df_bin_centroids |>
  dplyr::filter(n_h > 0)

trimesh <- trimesh |>
  dplyr::filter(from_count > 0,
                to_count > 0)

trimesh <- update_trimesh_index(trimesh)
\end{verbatim}

\subsubsection{Lifting the model into high dimensions}\label{lifting-the-model-into-high-dimensions}

The final step involves lifting the fitted \(2\text{-}D\) model into \(p\text{-}D\). This is done by modelling a point in \(p\text{-}D\) as the \(p\text{-}D\) mean of data points in the \(2\text{-}D\) centroid. This is performed using the \texttt{avg\_highd\_data()} function, which takes \(p\text{-}D\) data (\texttt{highd\_data}) and embedding data with their corresponding hexagonal bin IDs as inputs (\texttt{scaled\_nldr\_hexid}).

\begin{verbatim}
df_bin <- avg_highd_data(
  highd_data = scurve,
  scaled_nldr_hexid = hb_obj$data_hb_id
)

head(df_bin, 5)
\end{verbatim}

\begin{verbatim}
> # A tibble: 5 x 8
>       h     x1     x2    x3       x4       x5       x6       x7
>   <int>  <dbl>  <dbl> <dbl>    <dbl>    <dbl>    <dbl>    <dbl>
> 1    58 -0.371 1.91    1.92 -0.00827 0.00189   0.0170   0.00281
> 2    68  0.958 0.0854  1.29  0.00265 0.0171    0.0876  -0.00249
> 3    69  0.855 0.0917  1.51  0.00512 0.000325 -0.0130  -0.00395
> 4    70  0.731 0.129   1.68 -0.00433 0.00211  -0.0356  -0.00240
> 5    71  0.474 0.108   1.88 -0.00260 0.000128  0.00785  0.00170
\end{verbatim}

\subsection{Prediction}\label{prediction}

The \texttt{predict\_emb()} function is used to predict a point in a \(2\text{-}D\) embedding for a new \(p\text{-}D\) data point using the fitted model. This function is useful to predict \(2\text{-}D\) embedding irrespective of the NLDR technique.

In the prediction process, first, the nearest \(p\text{-}D\) model point is identified for the new \(p\text{-}D\) data point by computing \(p\text{-}D\) Euclidean distance. Then, the corresponding \(2\text{-}D\) bin centroid mapping for the identified \(p\text{-}D\) model point is determined. Finally, the coordinates of the identified \(2\text{-}D\) bin centroid is used as the predicted NLDR embedding for the new \(p\text{-}D\) data point.

To accelerate this process, the nearest-neighbor search is implemented in C++ using \texttt{Rcpp} via the internal function \texttt{compute\_highd\_dist()}.

\begin{verbatim}
predict_data <- predict_emb(
  highd_data = scurve,
  model_2d = df_bin_centroids,
  model_highd = df_bin
  )

head(predict_data, 5)
\end{verbatim}

\begin{verbatim}
> # A tibble: 5 x 4
>   pred_emb_1 pred_emb_2    ID pred_h
>        <dbl>      <dbl> <int>  <int>
> 1      0.252      0.901     1    427
> 2      0.692      0.545     2    287
> 3      0.780      0.393     3    226
> 4      0.164      0.952     4    446
> 5      0.193      1.00      5    468
\end{verbatim}

It is worth noting that while \texttt{predict\_emb()} provides a general approach that works across methods, some NLDR techniques have their own built-in prediction mechanisms. For example, UMAP \citep{tomasz2023} supports direct prediction of embeddings for new data once a model is fitted.

\subsection{Compute residuals and hexbin error (HBE)}\label{compute-residuals-and-hexbin-error-hbe}

Hexbin error (HBE) serves as a goodness-of-fit metric for evaluating the high-dimensional to \(2\text{-}D\) mapping. The \texttt{glance()} function computes these summary diagnostics by combining the fitted model returned by \texttt{fit\_highd\_model()} with the \(p\text{-}D\) data. After renaming the model output to avoid join conflicts, \texttt{predict\_emb()} is used to assign each observation to a hexagon in the \(2\text{-}D\) embedding. The resulting bin assignments are joined with both the model output and the \(p\text{-}D\) data, allowing squared and absolute differences between the true and modeled \(p\text{-}D\) coordinates to be computed for every dimension. Total absolute error (\texttt{Error}) and hexbin error (\texttt{HBE}, the root mean squared error) are then obtained using an efficient C++ implementation (\texttt{compute\_errors()}), and returned in a tidy tibble.

\begin{verbatim}
glance(
  x = scurve_model_obj,
  highd_data = scurve
)
\end{verbatim}

\begin{verbatim}
> # A tibble: 1 x 2
>   Error   HBE
>   <dbl> <dbl>
> 1  196. 0.116
\end{verbatim}

The \texttt{augment()} function provides point-level diagnostics by appending predictions and residuals to the original \(p\text{-}D\) data. Using the same prediction process as \texttt{glance()}, it computes dimension-wise residuals, squared errors, and absolute errors, along with two aggregate measures per observation: the total squared error (\texttt{row\_wise\_total\_error}) and the total absolute error (\texttt{row\_wise\_abs\_error}). The final output is a tibble containing IDs (\texttt{ID}), \(p\text{-}D\) data, predicted hexagons (\texttt{h}), modeled coordinates, and all residual diagnostics, with one row per observation.

\begin{verbatim}
model_error <- augment(
  x = scurve_model_obj,
  highd_data = scurve
)
\end{verbatim}

\subsection{Visualizations}\label{visualizations}

The package offers several \(2\text{-}D\) visualizations (Figure \ref{fig:geom-outputs-pdf}), including a full hexagonal grid, a hexagonal grid that matches the data, a full grid based on centroid triangulation, a centroid triangulation grid that aligns with the data, and a triangular mesh for any provided set of points.

The generated \(p\text{-}D\) model, overlaid with the data, can also be visualized using \texttt{show\_langevitour}. Additionally, it features a function for visualizing the \(2\text{-}D\) projection of the fitted model overlaid on the data, called \texttt{plot\_proj}.

Furthermore, there are two interactive plots, \texttt{show\_link\_plots} and \texttt{show\_error\_link\_plots}, which are designed to help diagnose the model. Each visualization can be generated using its respective function, as described in this section.

\subsubsection{Hexagonal grid}\label{hexagonal-grid}

The \texttt{geom\_hexgrid()} function introduces a custom \texttt{ggplot2} layer designed for visualizing hexagonal grid on a provided set of bin centroids.

To display the complete grid, users should supply all available bin centroids (Figure \ref{fig:geom-outputs-pdf}a).

\begin{verbatim}
full_hexgrid <- ggplot() +
  geom_hexgrid(
    data = hb_obj$centroids,
    aes(x = c_x, y = c_y)
    )
\end{verbatim}

If the goal is to plot only the subset of hexagons that correspond to bins containing data points, then only the centroids associated with those bins should be passed (Figure \ref{fig:geom-outputs-pdf}b).

\begin{verbatim}
data_hexgrid <- ggplot() +
  geom_hexgrid(
    data = df_bin_centroids,
    aes(x = c_x, y = c_y)
    )
\end{verbatim}

\subsubsection{Triangular mesh}\label{triangular-mesh}

The \texttt{geom\_trimesh()} function introduces a custom \texttt{ggplot2} layer designed for visualizing \(2\text{-}D\) wireframe on a provided set of bin centroids.

\begin{figure}[!ht]

{\centering \includegraphics[width=1\linewidth,alt={The figure consists of four 2-D panels illustrating hexagonal grids and triangulated meshes. Panel (a) shows a complete hexagonal grid covering a rectangular 2-D region, with all hexagons drawn regardless of whether data are present. Panel (b) shows a hexagonal grid restricted to bins that contain data points, with only those hexagons displayed and gaps where bins are empty. Panel (c) shows a full triangulated mesh constructed from centroids across the entire grid region, with centroids connected by straight line segments to form triangles. Panel (d) shows a triangulated mesh constructed only from centroids corresponding to non-empty bins, aligned with the data layout and leaving open areas where no centroids are present. In all panels, the horizontal and vertical axes represent continuous 2-D coordinates with bounded numeric ranges, and the grids and meshes are drawn with uniform line styles and colors.}]{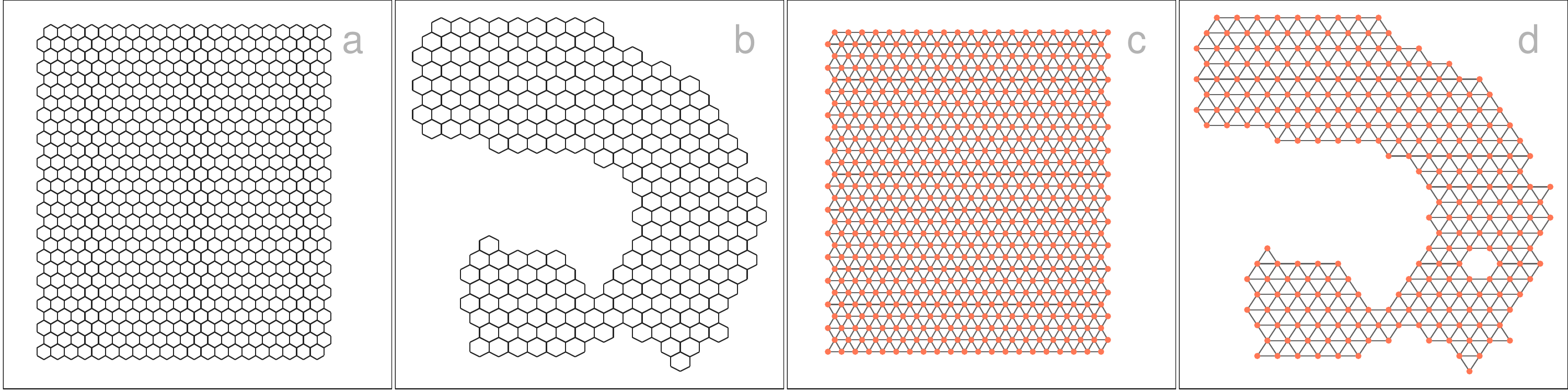}

}

\caption{The outputs of \texttt{geom\_hexgrid} and \texttt{geom\_trimesh} include: (a) a complete hexagonal grid, (b) a hexagonal grid that corresponds with the data, (c) a full grid based on centroid triangulation, and (d) a centroid triangulation grid that aligns with the data.}\label{fig:geom-outputs-pdf}
\end{figure}

To display the complete wireframe, users should supply all available bin centroids (Figure \ref{fig:geom-outputs-pdf}c).

\begin{verbatim}
full_triangulation_grid <- ggplot() +
  geom_trimesh(
    data = hb_obj$centroids,
    aes(x = c_x, y = c_y)
    )
\end{verbatim}

If the goal is to plot only the subset of hexagons that correspond to bins containing data points, then only the centroids associated with those bins should be passed (Figure \ref{fig:geom-outputs-pdf}d).

\begin{verbatim}
data_triangulation_grid <- ggplot() +
  geom_trimesh(
    data = df_bin_centroids,
    aes(x = c_x, y = c_y)
    )
\end{verbatim}

\subsubsection{\texorpdfstring{\(p\text{-}D\) model visualization}{p\textbackslash text\{-\}D model visualization}}\label{ptext-d-model-visualization}

\begin{figure}[!ht]
\includegraphics[width=1\linewidth,alt={The figure shows four panels of 2-D scatter plots (a1–a4), each displaying a different projection of the lifted high-dimensional wireframe model derived from the scurve UMAP layout. In each panel, the horizontal and vertical axes represent two projected coordinates with continuous numeric ranges. The wireframe model is drawn in black as a wireframe of connected line segments, and the scurve data points are overlaid in blue as individual markers. The relative positions and orientations of the wireframe and data points vary across panels as different projections are shown. In some regions, the wireframe extends into areas with few or no data points, leaving visible gaps between the model and the data. All panels have similar aspect ratios and use consistent color and symbol encodings to distinguish the model from the data.}]{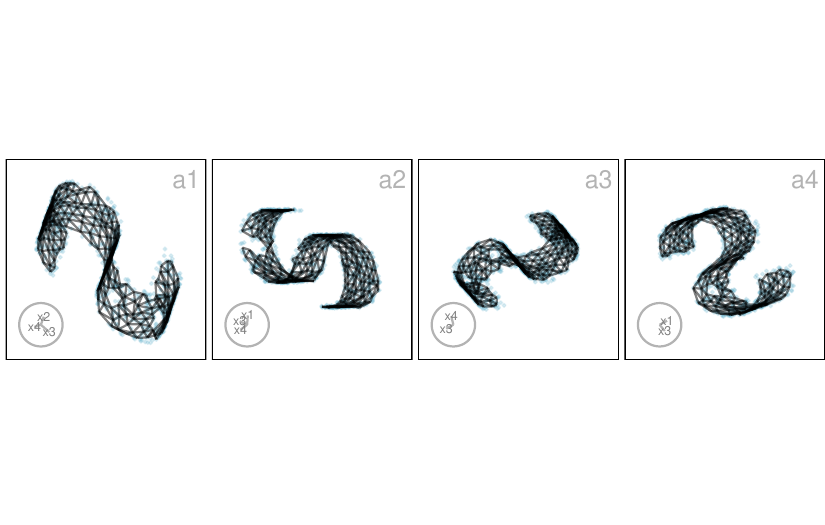} \caption{$2\text{-}D$ projections of the lifted high-dimensional wireframe model from the \texttt{scurve} UMAP layout. Each panel (a1–a4) shows the model (black) overlaid on \texttt{scurve} data (blue) in different projections. These views illustrate how the lifted wireframe model captures the structure of the \texttt{scurve} data. Regions with sparse or no data in the UMAP layout are also visible in the lifted model.}\label{fig:scurve-highd-model-pdf}
\end{figure}

To visualize how well the \(p\text{-}D\) model captures the underlying structure of the high-dimensional data, we provide a tour of the model in \(p\text{-}D\) using the \texttt{show\_langevitour()} function (Figure \ref{fig:scurve-highd-model-pdf}). This function renders a dynamic projection of both the high-dimensional data and the model using the \texttt{langevitour} R package \citep{paul2023}.

Before plotting, the data needs to be organized into a combined format through the \texttt{comb\_data\_model()} function. This function takes three inputs: \texttt{highd\_data} (the high-dimensional observations), \texttt{model\_highd} (high-dimensional summaries for each bin), and \texttt{model\_2d} (the hexagonal bin centroids of the model). It returns a tidy data frame combining both the data and the model.

In this structure, the \texttt{type} variable distinguishes between original observations (\texttt{data}) and the bin-averaged model representation (\texttt{model}).

\begin{verbatim}
df_exe <- comb_data_model(
  highd_data = scurve,
  model_highd = df_bin,
  model_2d = df_bin_centroids
  )
\end{verbatim}

The \texttt{show\_langevitour()} function then renders the visualization using the \texttt{langevitour} interface, displaying both types of points in a dynamic tour. The \texttt{edge\_data} input defines connections between neighboring bins (i.e., the hexagonal edges) to visualize the model's structure.

\begin{verbatim}
show_langevitour(
  point_data = df_exe,
  edge_data = trimesh
  )
\end{verbatim}

\begin{figure}[H]
\includegraphics[width=0.5\linewidth,alt={The figure shows several screenshots of 2-D projections of the lifted high-dimensional wireframe model derived from the scurve UMAP layout using detourr. In each screenshot, the horizontal and vertical axes represent projected coordinates with continuous numeric ranges. The wireframe model is displayed as a set of connected line segments, and the scurve data points are shown as individual markers overlaid on the wireframe. The orientation and shape of the wireframe and point cloud differ across screenshots as different projections are shown. In some views, parts of the wireframe extend into areas with few or no data points, leaving visible gaps between the model and the data. All screenshots use consistent color and line styles and have similar aspect ratios.}]{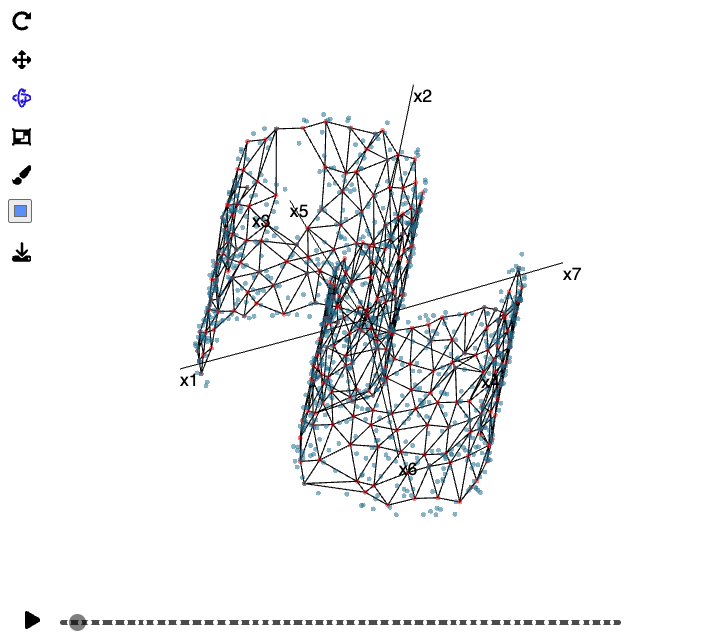} \includegraphics[width=0.5\linewidth,alt={The figure shows several screenshots of 2-D projections of the lifted high-dimensional wireframe model derived from the scurve UMAP layout using detourr. In each screenshot, the horizontal and vertical axes represent projected coordinates with continuous numeric ranges. The wireframe model is displayed as a set of connected line segments, and the scurve data points are shown as individual markers overlaid on the wireframe. The orientation and shape of the wireframe and point cloud differ across screenshots as different projections are shown. In some views, parts of the wireframe extend into areas with few or no data points, leaving visible gaps between the model and the data. All screenshots use consistent color and line styles and have similar aspect ratios.}]{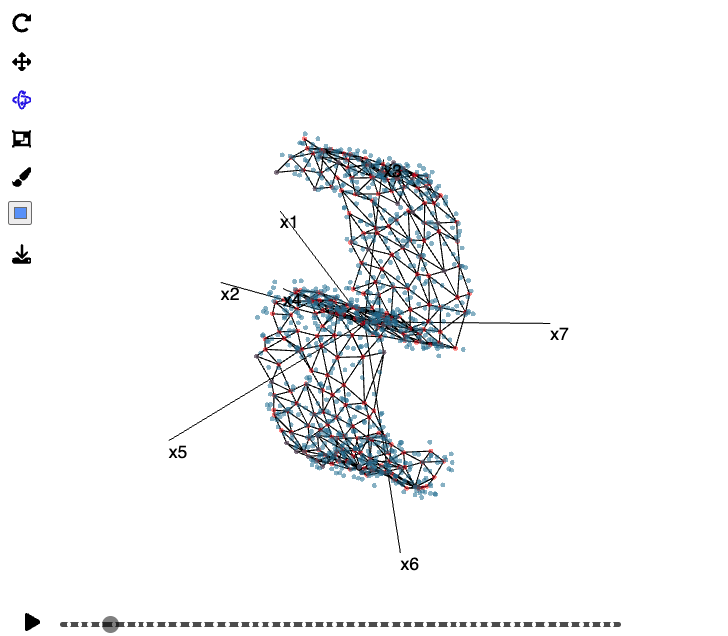} \caption{Screenshots of the lifted high-dimensional wireframe model from the \texttt{scurve} UMAP layout using \texttt{detourr}. Regions with sparse or no data in the UMAP layout are also visible in the lifted model.}\label{fig:scurve-highd-model-detourrpdf}
\end{figure}

As an alternative to \texttt{langevitour}, users can explore the fitted \(p\text{-}D\) model using the \texttt{detourr} \citep{casper2025} (Figure \ref{fig:scurve-highd-model-detourrpdf}). The combined data object from \texttt{comb\_data\_model()} can be passed directly to the \texttt{detour()} function, where \texttt{tour\_aes()} defines the projection variables and color mapping. The visualization is rendered using \texttt{show\_scatter()}, which can display both data points and the model's structural edges via the \texttt{edges} argument.

\begin{verbatim}
detour(
  df_exe,
  tour_aes(
    projection = starts_with("x"),
    colour = type
  )
) |>
  tour_path(grand_tour(2),
                    max_bases=50, fps = 60) |>
  show_scatter(axes = TRUE, size = 1.5, alpha = 0.5,
               edges = as.matrix(trimesh[, c("from_reindexed", "to_reindexed")]),
               palette = c("#66B2CC", "#FF7755"),
               width = "600px", height = "600px")
\end{verbatim}

In the resulting interactive visualization, blue points represent the high-dimensional data, orange points represent the model centroids from each bin, and the lines between model points reflect the \(2\text{-}D\) wireframe structure mapped to high-dimensional space.

\subsubsection{Linked plots}\label{linked-plots}

Two types of interactively linked plots can be generated to assess the model fits everywhere, or better in some subspaces, or completely mismatch the data. The plots are linked using \texttt{crosstalk}, which allows interactive brushing: selecting or brushing points in one plot automatically highlights the corresponding points in the other linked views.

\begin{figure}[!ht]
\includegraphics[width=1\linewidth,alt={The figure consists of four linked panels arranged in two rows and two columns. Panels (a1) and (b1) show 2-D scatter plots of a UMAP layout, with the horizontal and vertical axes representing the two UMAP coordinates on continuous numeric scales. Panels (a2) and (b2) show corresponding 2-D projections from a tour of the 7-D data and model, where each axis represents a linear combination of the original seven dimensions. Data points are plotted as individual markers, and selected subsets are highlighted in purple while unselected points appear in a contrasting color. In the top row, a set of points highlighted in the lower region of the UMAP layout in (a1) is linked to highlighted points in the tour view in (a2). In the bottom row, a different highlighted region in the UMAP layout in (b1) is linked to a different set of highlighted points in the tour view in (b2). The linked selections update consistently across panels, showing how the same subsets of observations appear in the UMAP layout and in different projections of the high-dimensional space.}]{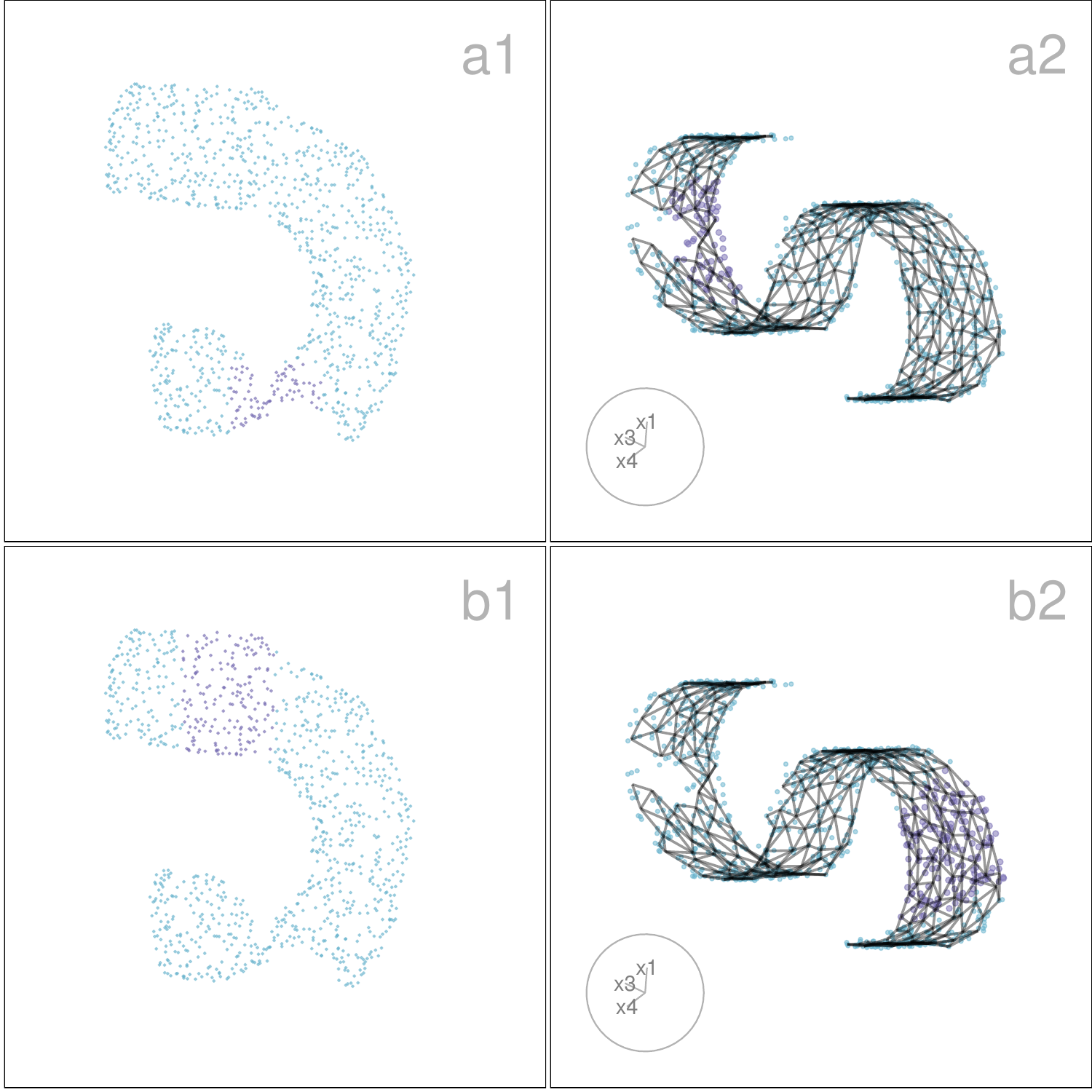} \caption{Exploring the correspondence between UMAP layout and \texttt{scurve} structure in $7\text{-}D$. Two sets of plots are interactively linked: UMAP layout (a1, b1) and projection of $7\text{-}D$ model and data (a2, b2). The purple points indicate the selected subsets, which differ between rows. In (a1), the lower bridge of the \texttt{scurve} is highlighted, which corresponds in (a2) to points spanning across both arms of the high-dimensional structure. In (b1), a different region near the upper arm of the \texttt{scurve} is selected, and in (b2) these points map onto one side of the curved manifold in $7\text{-}D$ projection. While the UMAP layout suggests distinct local clusters, the linked tour views reveal how these selections trace continuous structures in the $7\text{-}D$ space, highlighting distortions introduced by UMAP.}\label{fig:scurve-nldrlink-pdf}
\end{figure}

First, the function \texttt{show\_link\_plots()} provides linking a \(2\text{-}D\) NLDR layout and a tour (via \texttt{langevitour}) of the model overlaid the data (Figure \ref{fig:scurve-nldrlink-pdf}).

The \texttt{point\_data} for \texttt{show\_link\_plots()} can be prepared using the \texttt{comb\_all\_data\_model()} function. This function combines the high-dimensional data (\texttt{highd\_data}), the NLDR data (\texttt{nldr\_data}), and the bin-averaged high-dimensional model representation (\texttt{model\_highd}) aligned to the \(2\text{-}D\) bin layout (\texttt{model\_2d}). This combined dataset includes both the original observations and the bin-level model averages, labeled with a \texttt{type} variable for distinguishing between them. Also, the \texttt{show\_link\_plots()} function takes \texttt{edge\_data}, which defines connections between neighboring bins.

\begin{verbatim}
df_exe <- comb_all_data_model(
  highd_data = scurve,
  nldr_data = scurve_umap,
  model_highd = df_bin,
  model_2d = df_bin_centroids
  )
\end{verbatim}

\begin{verbatim}
nldrdt_link <- show_link_plots(
  point_data = df_exe,
  edge_data = trimesh,
  point_colour = clr_choice
  )

nldrdt_link <- crosstalk::bscols(
    htmltools::div(
        style = "display: grid; grid-template-columns: 1fr 1fr;
    gap: 0px; align-items: start; justify-items: center; margin: 0; padding: 0;
    height: 380px; width: 500px",
        nldrdt_link
    ),
    device = "xs"
)

class(nldrdt_link) <- c(class(nldrdt_link), "htmlwidget")

nldrdt_link
\end{verbatim}

\begin{figure}[H]
\includegraphics[width=1\linewidth,alt={The figure consists of six panels arranged in two rows and three columns, showing residuals and their correspondence between UMAP and high-dimensional projections of the scurve data. In each row, the left panel (a1, b1) is a univariate plot of model residuals, with the horizontal axis representing residual values and the vertical axis representing frequency or density; points selected for highlighting appear in purple. The middle panel (a2, b2) shows a 2-D UMAP layout, with horizontal and vertical axes representing the two UMAP coordinates; the selected points from the left panel are highlighted in purple, while unselected points appear in a contrasting color. The right panel (a3, b3) shows a 2-D projection of the 7-D data and model using a tour view, where axes represent linear combinations of the original dimensions; highlighted points correspond to those selected in the residuals and UMAP panels, while unselected points are shown in a contrasting color. Across the panels, the purple highlights differ between the top and bottom rows, showing different subsets of points. All panels maintain roughly square aspect ratios and use consistent visual encodings for selection and non-selection.}]{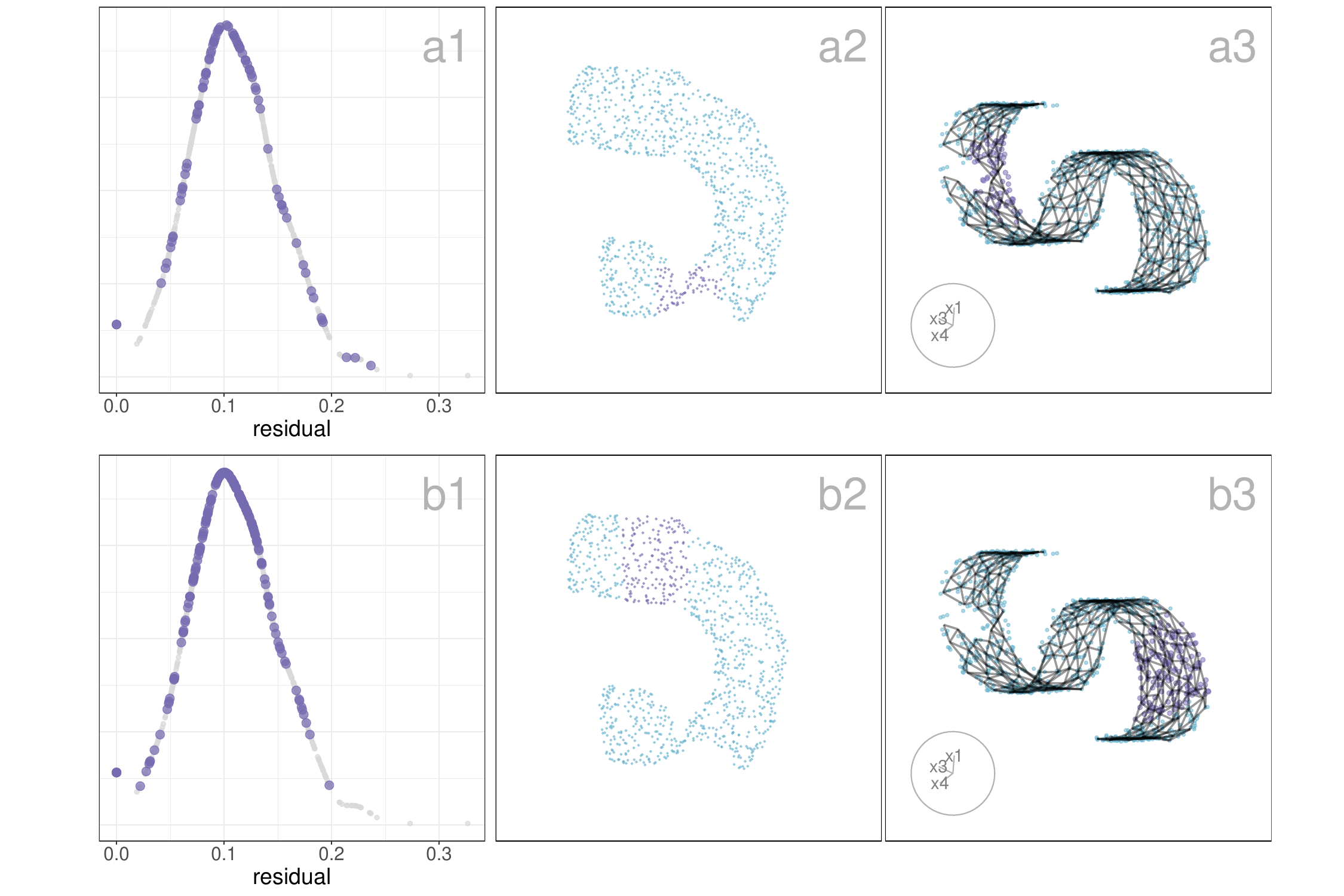} \caption{Exploring residuals in relation to UMAP layouts using a $7\text{-}D$ \texttt{scurve} model. Three views are linked: distribution of residuals (a1, b1), UMAP layout (a2, b2), and projection of the $7\text{-}D$ model with data (a3, b3). The purple points highlight selected subsets of the data, which differ across rows. In the top row (a1–a3), points with higher residuals (a1) are selected, corresponding to the sparse bridging region in the UMAP layout (a2) and the less dense end of the \texttt{scurve} in the high-dimensional projection (a3). In the bottom row (b1–b3), points with lower residuals (b1) are highlighted, which map to one side of the dense region in the NLDR layout (b2) and to a thicker band of the \texttt{scurve} in the projection (b3). This comparison illustrates how residuals can help diagnose distortions in UMAP, with high-residual points often concentrated in sparse or stretched regions of the structure.}\label{fig:scurve-linkerror-pdf}
\end{figure}

The function \texttt{show\_error\_link\_plots()} generates three side-by-side, interactively linked plots; a error distribution, a \(2\text{-}D\) NLDR layout, and a tour (via \texttt{langevitour}) of the model overlaid the data (Figure \ref{fig:scurve-linkerror-pdf}). The function takes the output from \texttt{comb\_all\_data\_model\_error()} (\texttt{point\_data}) and \texttt{edge\_data} which defines connections between neighboring bins.

The \texttt{point\_data} can be generated using the \texttt{comb\_all\_data\_model\_error()} function. The function requires several arguments: points data which contain high-dimensional data (\texttt{highd\_data}), NLDR data (\texttt{nldr\_data}), high-dimensional model data (\texttt{model\_highd}), \(2\text{-}D\) model data (\texttt{model\_2d}), and model error (\texttt{error\_data}). This combined dataset includes both the original observations and the bin-level model averages, labeled with a \texttt{type} variable for distinguishing between them.

\begin{verbatim}
df_exe <- comb_all_data_model_error(
  highd_data = scurve,
  nldr_data = scurve_umap,
  model_highd = df_bin,
  model_2d = df_bin_centroids,
  error_data = model_error
  )
\end{verbatim}

\begin{verbatim}
errornldrdt_link <- show_error_link_plots(
  point_data = df_exe,
  edge_data = trimesh,
  point_colour = clr_choice
)

class(errornldrdt_link) <- c(class(errornldrdt_link), "htmlwidget")

errornldrdt_link
\end{verbatim}

In the implementation examples, points are shown without cluster-based coloring unless explicitly stated. When points are colored by cluster, linked plot functionality is currently only partially supported: selections made in the \texttt{langevitour} controls are not reflected in the corresponding highlights in the interactive \(2\text{-}D\) layout. As a result, cluster-specific exploration must be carried out separately in the tour view and the interactive \(2\text{-}D\) layout. In addition to \texttt{langevitour}, linked plots can also be constructed using tour views generated with the \texttt{detourr} package. (see \href{https://jayanilakshika.github.io/quollr/articles/quollr8linkeddetourr.html}{vignette: 8. Linked plots with \texttt{detourr}} for details.)

\section{Application}\label{application}

Single-cell RNA sequencing (scRNA-seq) is a popular and powerful technology that allows you to profile the whole transcriptome of a large number of individual cells \citep{andrews2021}.

Clustering of single-cell data is used to identify groups of cells with similar expression profiles. NLDR often used to summarize the discovered clusters, and help to understand the results. The purpose of this example is to \emph{illustrate how to use our method to help decide on an appropriate NLDR layout that accurately represents the data}.

Limb muscle cells of mice in \citet{tabula2018} are examined. There are \(1067\) single cells, with \(14997\) gene expressions. Following their pre-processing, different NLDR methods were performed using ten principal components. Figure \ref{fig:limb-hbe} (a) is the reproduction of the published plot. This was generated using tSNE with \(\text{perplexity}=30\), the default hyper-parameters. The question is whether this accurately represents the cluster structure in the data. Note that the cluster variable is not used to produce the \(2\text{-}D\) layout.

We illustrate how to use \texttt{quollr} to assess whether this is a reasonable layout. Figure \ref{fig:limb-hbe} shows five alternative layouts, and the HBE plot summarizing the resulting model fits. Layout b is produce by UMAP (\(\text{neighbors}=5, ~\text{minimum distance}=0.1\)); layout c was produced by PHATE (\(\text{knn}=5\)); layout d was produced by TriMAP (\(\text{number of inliers}=12,~\text{outliers}=4,~\text{random}=3\)); layout e was produced by by PaCMAP (\(\text{neighbors}=10,~\text{init = "random"}, ~\text{MN-ratio}=0.5,~\text{FP-ratio}=2\)); layout f was produced by tSNE (\(\text{perplexity}=15\)).

The HBE plot indicates that the two tSNE layouts outperform all the other methods across a range of binwidths, but that the result with perplexity of \(15\) outperforms the other. There are small visual difference between the two layouts. Both support that there are \(5-6\) clusters. Layout a has slightly more space between clusters. Layout d three small clusters at the top whereas layout f has only two, and another smaller one at the bottom.

\begin{verbatim}
design <- gen_design(n_right = 6, ncol_right = 2)

plot_hbe_layouts(plots = list(error_plot_limb, nldr1,
                             nldr2, nldr3, nldr4,
                             nldr5, nldr6), design = design)
\end{verbatim}

\begin{figure}[!ht]
\includegraphics[width=1\linewidth,alt={A multi-panel figure shows five alternative 2D embeddings of the same high-dimensional data, arranged side by side (panels b–f). Each panel is a scatterplot with two unlabeled embedding axes (dimension 1 on the horizontal axis and dimension 2 on the vertical axis, each spanning a similar, roughly symmetric range around zero). Points represent the same set of observations in every panel, plotted with identical color or marker scheme so that corresponding observations can be visually compared across layouts. Panel b shows a UMAP layout (neighbors = 5, minimum distance = 0.1), panel c a PHATE layout (knn = 5), panel d a TriMAP layout (inliers = 12, outliers = 4, random = 3), panel e a PaCMAP layout (neighbors = 10, init = random, MN-ratio = 0.5, FP-ratio = 2), and panel f a tSNE layout (perplexity = 15). Across panels, similar clusters of points appear but are arranged differently in space: some layouts show more clearly separated compact clusters, while others show overlapping or elongated groups, reflecting how each method balances local versus global structure. The accompanying HBE plot (typically a separate panel) summarizes model fit quality for each layout, with one value per method indicating that some embeddings provide a more faithful representation of the original data structure than others.}]{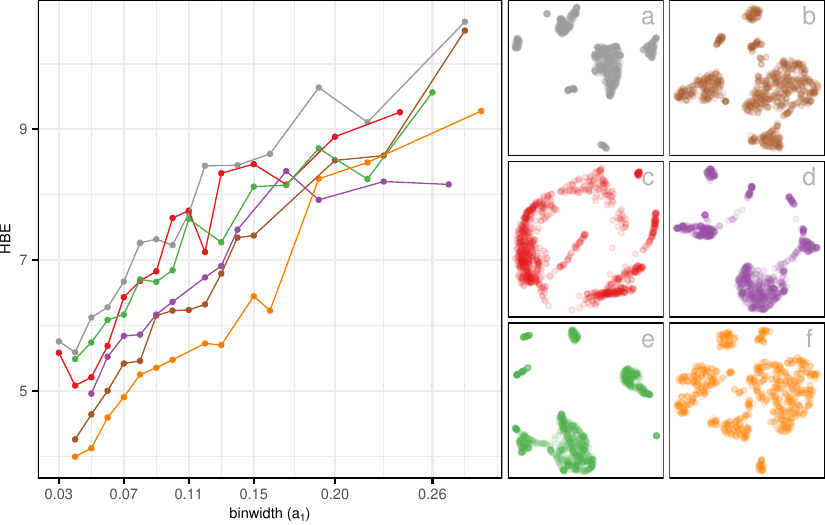} \caption{Assessing which of the 6 NLDR layouts on the limb muscle data is the better representation using HBE for varying binwidth ($a_1$). Color used for the lines and points in the left plot and in the scatterplots represents NLDR layout (a-f). Layout d performs well at large binwidth (where the binwidth is not enough to capture the data structure) and poorly as the bin width decreases. Layout f is the best choice.\label{fig:limb-hbe}}\label{fig:limb-hbe}
\end{figure}

\begin{figure}[!ht]
\includegraphics[width=1\linewidth,alt={The figure consists of six panels arranged in two rows and three columns, showing representative views of two NLDR layouts for the Limb muscle dataset (n=1067). In the top row (a1–a3), the left panel (a1) shows the published 2-D NLDR embedding with points colored by muscle group and overlaid with triangulated hexagon centroids. The middle (a2) and right (a3) panels show two different 2-D projections of the corresponding 10-D data and fitted model, with points plotted as individual markers and the same triangular mesh connecting centroids. In the bottom row (f1–f3), the left panel (f1) shows the 2-D NLDR layout selected using the HBE plot, again with colored points and triangulated centroids, while panels (f2) and (f3) show two different 2-D projections of the associated 10-D structure with the same mesh displayed. Across all panels, axes represent continuous numeric coordinates, color encodes muscle group, and the triangular mesh is used consistently to show correspondence between the low-dimensional layouts and the high-dimensional projections.}]{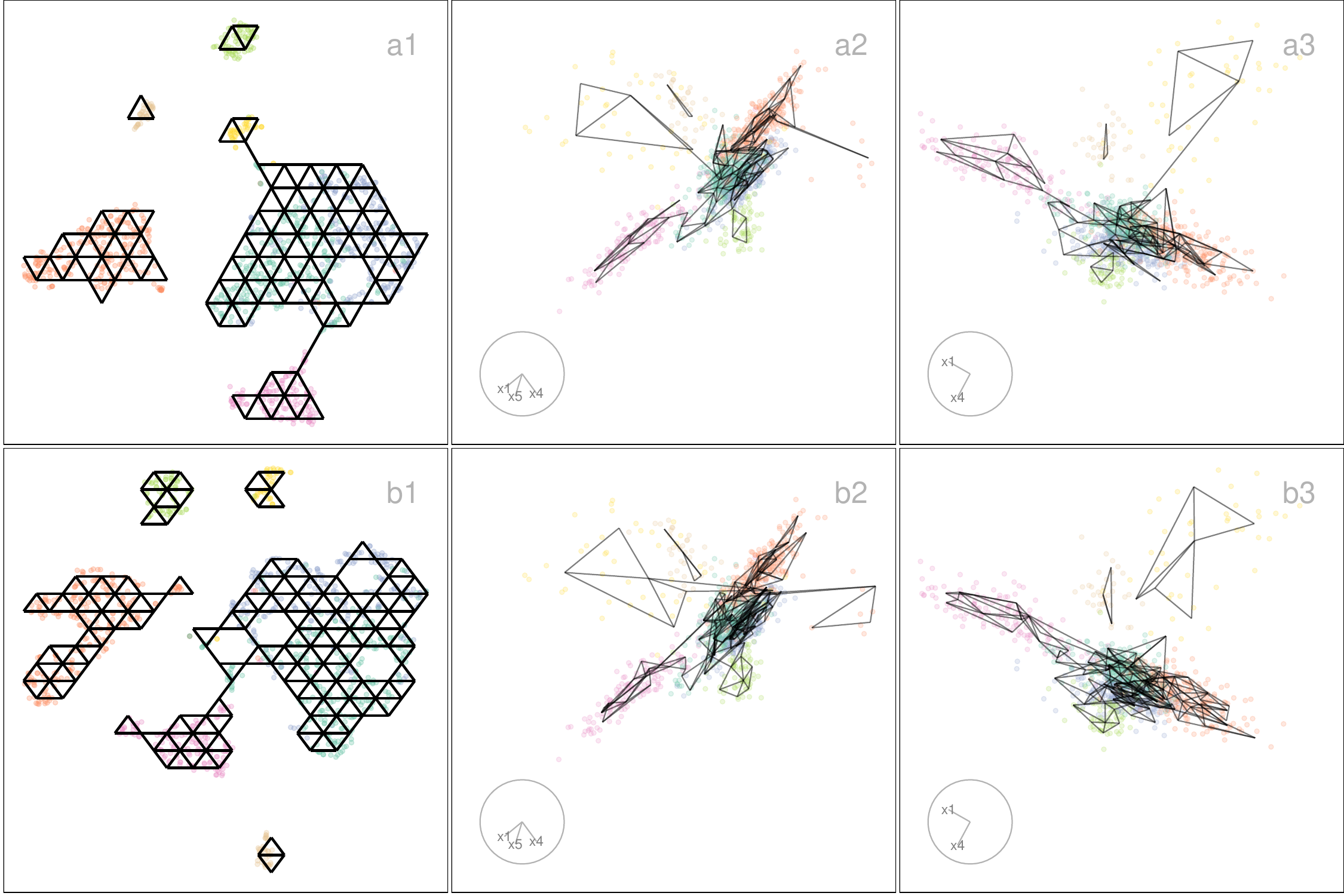} \caption{Representative views of two selected NLDR layouts for the Limb muscle dataset ($n=1067$), shown row-wise. The top row (a1–a3) corresponds to the published $2\text{-}D$ layout (Figure \ref{fig:limb-hbe}a), and the bottom row (f1–f3) corresponds to the $2\text{-}D$ layout selected (Figure \ref{fig:limb-hbe}f) selected using the HBE plot. In each row, the left panel (a1, f1) shows the NLDR embedding with points colored by muscle group and overlaid with triangulated hexagon centroids. The middle (a2, f2) and right (a3, f3) panels show two different $2\text{-}D$ projections of the fitted model and data in the original $10\text{-}D$ space, with the same triangular mesh displayed. Together, these panels summarize how the low-dimensional layouts relate to the underlying high-dimensional structure across different viewing directions.}\label{fig:model-limb}
\end{figure}

Figure \ref{fig:model-limb} show how to examine the resulting models of the layouts overlaid on the data in high dimensions. Point color represents the cluster reported in the published paper. In each case the best model is produced using \(\text{binwidth}=0.06\). A binwidth of \(0.06\) is used because it is small enough to show local structure and differences between clusters, but large enough to avoid breaking the layout into too many small pieces that would make the plots hard to interpret. The plots in these figures are best understand using small steps.

\begin{enumerate}
\def\labelenumi{\arabic{enumi}.}
\item
  Examine the model in the data space for each, by looking at the tour views. In each case, the clustering doesn't quite match the separations in the data, and both models help see this. For example, the orange cluster (\(1\)) should probably be split into more than one cluster because both models show large stretched lines connecting a small group far from the remaining orange points.
\item
  Because \(6\) clusters are hard to examine together, use the menu to select just one cluster to view at at time. Selecting just cluster \(1\) might help you see the explanation above, that a small group is quite separate from the main group. This suggests both the layouts and the clustering that groups these together might be wrong. Now change to focus on cluster \(5\) (yellow). This group is a fairly large sparse cluster but it is separated from the other points. Both layouts are right in separating these points from the others but the fit for layout f is slightly better.
\item
  To examine where the layouts differ, examine clusters \(4\) (green) and \(6\) (blue), by selecting just these two. Layout a has them close, but layout f has them far apart. (It might also help to include cluster \(5\) here because layout a has this group close to the cluster \(4\) also). In the tour view, we can see that they three clusters are separated clusters in different directions away from the most of the other points. They are both correct in this. It may have been better to place them all in different corners of the layout, but they have preserved the most important aspect that they are separated clusters. That they are all close together in layout a could be incorrectly interpreted as close in high-dimensions though.
\end{enumerate}

\section{Discussion}\label{discussion}

The \texttt{quollr} package introduces a new framework for interpreting NLDR outputs by fitting a geometric wireframe model in \(2\text{-}D\) and lifting it into high-dimensional space. This lifted model provides a direct way to assess how well a \(2\text{-}D\) layout, produced by methods such as tSNE, UMAP, PHATE, TriMAP, or PaCMAP, preserves the structure of the original high-dimensional data. The approach offers both numerical and visual diagnostics to support the selection of NLDR methods and tuning hyper-parameters that produce the most accurate \(2\text{-}D\) representations.

In contrast to the common practice of visually inspecting scatterplots for clusters or patterns, \texttt{quollr} provides a quantitative route for evaluation. It enables the computation of HBE and residuals between the original high-dimensional data and the lifted model, offering interpretable diagnostics. These diagnostics are complemented by interactive linked plots and high-dimensional dynamic visualizations using the \texttt{langevitour} package, allowing users to inspect where the model fits well and where it does not.

To support efficient computation, particularly for large-scale datasets, several core functions in \texttt{quollr} are implemented in C++ using \texttt{Rcpp} and \texttt{RcppArmadillo}. These include functions for computing Euclidean distances in high-dimensional and \(2\text{-}D\) space, identifying nearest centroids, calculating residual errors, and generating polygonal coordinates of hexagons. For instance, \texttt{compute\_highd\_dist()} accelerates nearest neighbor lookup in high-dimensional space, \texttt{compute\_errors()} calculates HBE and total absolute error efficiently, and \texttt{calc\_2d\_dist\_cpp()} speeds up distance calculations in \(2\text{-}D\). Additionally, \texttt{gen\_hex\_coord\_cpp()} constructs the coordinates for hexagonal bins based on their centroids with minimal overhead. These optimizations result in substantial performance gains compared to native R implementations, making the package responsive even when used in interactive contexts or on large datasets such as single-cell transcriptomic profiles.

The modular structure of the package is designed to support both flexibility and reproducibility. Users can access individual functions to control each step of the pipeline such as scaling, binning, and triangulation or use the main function \texttt{fit\_highd\_model()} for end-to-end model construction. The diagnostics can be used not only to compare NLDR methods but also to tune binning parameters, assess layout stability, and detect local distortions in the embedding.

There are several avenues for future development. While hexagonal binning provides a regular structure conducive to modeling, alternative spatial discretizations (e.g., adaptive binning or density-aware tessellations) could be explored to better capture varying data densities. Expanding support for additional distance metrics in the lifting and prediction steps may improve performance across different domains. Additionally, statistical inference tools could be introduced to assess the stability and robustness of the fitted model, which would enhance interpretability and confidence in the outcomes. A potential extension of the current implementation would be to synchronize cluster selections between the tour view and the linked \(2\text{-}D\) layout, enabling more direct cluster-specific comparisons across views.

\section{Acknowledgements}\label{acknowledgements}

The source code for reproducing this paper can be found at: \href{https://github.com/JayaniLakshika/paper-quollr}{github.com/JayaniLakshika/paper-quollr}. This article is created using \CRANpkg{knitr} \citep{yihui2015} and \CRANpkg{rmarkdown} \citep{yihui2018} in R with the \texttt{rjtools::rjournal\_article} template. These \texttt{R} packages were used for this work: \texttt{cli} \citep{gabor2025}, \texttt{dplyr} \citep{hadley2023}, \texttt{ggplot2} \citep{hadley2016}, \texttt{interp} (\textgreater= 1.1-6) \citep{albrecht2024}, \texttt{langevitour} \citep{paul2023}, \texttt{detourr} \citep{casper2025}, \texttt{proxy}\citep{david2022}, \texttt{stats} \citep{core2025}, \texttt{tibble} \citep{kirill2023}, \texttt{tidyselect} \citep{lionel2024}, \texttt{crosstalk} \citep{joe2023}, \texttt{plotly} \citep{chapman2020}, \texttt{htmltools} \citep{joe2024}, \texttt{kableExtra} \citep{hao2024}, \texttt{patchwork} \citep{thomas2024}, and \texttt{readr} \citep{hadley2024}.

\bibliography{paper-quollr.bib}

\begin{thebibliography}{28}
\providecommand{\natexlab}[1]{#1}
\providecommand{\url}[1]{\texttt{#1}}
\expandafter\ifx\csname urlstyle\endcsname\relax
  \providecommand{\doi}[1]{doi: #1}\else
  \providecommand{\doi}{doi: \begingroup \urlstyle{rm}\Url}\fi

\bibitem[Amid and Warmuth(2019)]{amid2019}
E.~Amid and M.~K. Warmuth.
\newblock Trimap: Large-scale dimensionality reduction using triplets.
\newblock \emph{ArXiv}, abs/1910.00204, 2019.
\newblock URL \url{https://api.semanticscholar.org/CorpusID:203610264}.

\bibitem[Andrews et~al.(2021)Andrews, Kiselev, McCarthy, and Hemberg]{andrews2021}
T.~S. Andrews, V.~Y. Kiselev, D.~McCarthy, and M.~Hemberg.
\newblock Tutorial: guidelines for the computational analysis of single-cell rna sequencing data.
\newblock \emph{Nature Protocols}, 16\penalty0 (1):\penalty0 1--9, 2021.
\newblock URL \url{https://doi.org/10.1038/s41596-020-00409-w}.

\bibitem[Cheng and Sievert(2023)]{joe2023}
J.~Cheng and C.~Sievert.
\newblock \emph{crosstalk: Inter-Widget Interactivity for HTML Widgets}, 2023.
\newblock URL \url{https://CRAN.R-project.org/package=crosstalk}.
\newblock R package version 1.2.1.

\bibitem[Cheng et~al.(2024)Cheng, Sievert, Schloerke, Chang, Xie, and Allen]{joe2024}
J.~Cheng, C.~Sievert, B.~Schloerke, W.~Chang, Y.~Xie, and J.~Allen.
\newblock \emph{htmltools: Tools for HTML}, 2024.
\newblock URL \url{https://CRAN.R-project.org/package=htmltools}.
\newblock R package version 0.5.8.1.

\bibitem[Cs\'ardi(2025)]{gabor2025}
G.~Cs\'ardi.
\newblock \emph{cli: Helpers for Developing Command Line Interfaces}, 2025.
\newblock URL \url{https://CRAN.R-project.org/package=cli}.
\newblock R package version 3.6.4.

\bibitem[et~al.(2018)]{tabula2018}
T.~M.~C. et~al.
\newblock Single-cell transcriptomics of 20 mouse organs creates a tabula muris.
\newblock \emph{Nature}, 562\penalty0 (7727):\penalty0 367--372, 2018.
\newblock \doi{10.1038/s41586-018-0590-4}.

\bibitem[Gamage et~al.(2025)Gamage, Cook, Harrison, Lydeamore, and Talagala]{gamage2025c}
J.~P. Gamage, D.~Cook, P.~Harrison, M.~Lydeamore, and T.~S. Talagala.
\newblock Choosing better nldr layouts by evaluating the model in the high-dimensional data space, 2025.
\newblock URL \url{https://arxiv.org/abs/2506.22051}.

\bibitem[Gebhardt et~al.(2024)Gebhardt, Bivand, and Sinclair]{albrecht2024}
A.~Gebhardt, R.~Bivand, and D.~Sinclair.
\newblock \emph{interp: Interpolation Methods}, 2024.
\newblock URL \url{https://CRAN.R-project.org/package=interp}.
\newblock R package version 1.1-6.

\bibitem[Harrison(2023)]{paul2023}
P.~Harrison.
\newblock langevitour: Smooth interactive touring of high dimensions, demonstrated with scrna-seq data.
\newblock \emph{The R Journal}, 15\penalty0 (2):\penalty0 206--219, 2023.
\newblock \doi{10.32614/RJ-2023-046}.

\bibitem[Hart and Wang(2025)]{casper2025}
C.~Hart and E.~Wang.
\newblock \emph{detourr: Portable and Performant Tour Animations}, 2025.
\newblock URL \url{https://CRAN.R-project.org/package=detourr}.
\newblock R package version 0.2.0.

\bibitem[Henry and Wickham(2024)]{lionel2024}
L.~Henry and H.~Wickham.
\newblock \emph{tidyselect: Select from a Set of Strings}, 2024.
\newblock URL \url{https://CRAN.R-project.org/package=tidyselect}.
\newblock R package version 1.2.1.

\bibitem[Konopka(2023)]{tomasz2023}
T.~Konopka.
\newblock \emph{umap: Uniform Manifold Approximation and Projection}, 2023.
\newblock URL \url{https://CRAN.R-project.org/package=umap}.
\newblock R package version 0.2.10.0.

\bibitem[Lee and Schachter(1980)]{lee1980}
D.~T. Lee and B.~J. Schachter.
\newblock Two algorithms for constructing a delaunay triangulation.
\newblock \emph{International Journal of Computer \& Information Sciences}, 9\penalty0 (3):\penalty0 219--242, 1980.
\newblock URL \url{https://doi.org/10.1007/BF00977785}.

\bibitem[Maaten and Hinton(2008)]{laurens2008}
L.~V.~D. Maaten and G.~E. Hinton.
\newblock Visualizing data using t-sne.
\newblock \emph{Journal of Machine Learning Research}, 9:\penalty0 2579--2605, 2008.

\bibitem[McInnes et~al.(2018)McInnes, Healy, Saul, and Gro\ss{}berger]{leland2018}
L.~McInnes, J.~Healy, N.~Saul, and L.~Gro\ss{}berger.
\newblock {UMAP}: Uniform manifold approximation and projection.
\newblock \emph{Journal of Open Source Software}, 3\penalty0 (29):\penalty0 861, 2018.
\newblock URL \url{https://doi.org/10.21105/joss.00861}.

\bibitem[Meyer and Buchta(2022)]{david2022}
D.~Meyer and C.~Buchta.
\newblock \emph{proxy: Distance and Similarity Measures}, 2022.
\newblock URL \url{https://CRAN.R-project.org/package=proxy}.
\newblock R package version 0.4-27.

\bibitem[Moon et~al.(2019)Moon, van Dijk, Wang, Gigante, Burkhardt, Chen, Yim, van~den Elzen, Hirn, Coifman, Ivanova, Wolf, and Krishnaswamy]{moon2019}
K.~R. Moon, D.~van Dijk, Z.~Wang, S.~A. Gigante, D.~B. Burkhardt, W.~S. Chen, K.~Yim, A.~van~den Elzen, M.~J. Hirn, R.~R. Coifman, N.~B. Ivanova, G.~Wolf, and S.~Krishnaswamy.
\newblock Visualizing structure and transitions in high-dimensional biological data.
\newblock \emph{Nature Biotechnology}, 37:\penalty0 1482--1492, 2019.
\newblock \doi{10.1038/s41587-019-0336-3}.

\bibitem[M\"uller and Wickham(2023)]{kirill2023}
K.~M\"uller and H.~Wickham.
\newblock \emph{tibble: Simple Data Frames}, 2023.
\newblock URL \url{https://CRAN.R-project.org/package=tibble}.
\newblock R package version 3.2.1.

\bibitem[Pedersen(2024)]{thomas2024}
T.~L. Pedersen.
\newblock \emph{patchwork: The Composer of Plots}, 2024.
\newblock URL \url{https://CRAN.R-project.org/package=patchwork}.
\newblock R package version 1.3.0.

\bibitem[{R Core Team}(2025)]{core2025}
{R Core Team}.
\newblock \emph{R: A Language and Environment for Statistical Computing}, 2025.
\newblock URL \url{https://www.R-project.org/}.

\bibitem[Sievert(2020)]{chapman2020}
C.~Sievert.
\newblock \emph{Interactive Web-Based Data Visualization with R, plotly, and shiny}.
\newblock Chapman and Hall/CRC, 2020.
\newblock URL \url{https://plotly-r.com}.

\bibitem[Wang et~al.(2021)Wang, Huang, Rudin, and Shaposhnik]{yingfan2021}
Y.~Wang, H.~Huang, C.~Rudin, and Y.~Shaposhnik.
\newblock Understanding how dimension reduction tools work: An empirical approach to deciphering t-sne, umap, trimap, and pacmap for data visualization.
\newblock \emph{Journal of Machine Learning Research}, 22\penalty0 (201):\penalty0 1--73, 2021.
\newblock URL \url{http://jmlr.org/papers/v22/20-1061.html}.

\bibitem[Wickham(2016)]{hadley2016}
H.~Wickham.
\newblock \emph{ggplot2: Elegant Graphics for Data Analysis}.
\newblock Springer-Verlag New York, 2016.
\newblock URL \url{https://ggplot2.tidyverse.org}.

\bibitem[Wickham(2023)]{hadley2023}
H.~Wickham.
\newblock \emph{conflicted: An Alternative Conflict Resolution Strategy}, 2023.
\newblock URL \url{https://CRAN.R-project.org/package=conflicted}.
\newblock R package version 1.2.0.

\bibitem[Wickham et~al.(2024)Wickham, Hester, and Bryan]{hadley2024}
H.~Wickham, J.~Hester, and J.~Bryan.
\newblock \emph{readr: Read Rectangular Text Data}, 2024.
\newblock URL \url{https://CRAN.R-project.org/package=readr}.
\newblock R package version 2.1.5.

\bibitem[Xie(2015)]{yihui2015}
Y.~Xie.
\newblock \emph{Dynamic Documents with {R} and knitr}.
\newblock Chapman and Hall/CRC, 2nd edition, 2015.
\newblock URL \url{https://yihui.name/knitr/}.

\bibitem[Xie et~al.(2018)Xie, Allaire, and Grolemund]{yihui2018}
Y.~Xie, J.~Allaire, and G.~Grolemund.
\newblock \emph{{R} Markdown: The Definitive Guide}.
\newblock Chapman and Hall/CRC, 2018.
\newblock URL \url{https://bookdown.org/yihui/rmarkdown}.

\bibitem[Zhu(2024)]{hao2024}
H.~Zhu.
\newblock \emph{kableExtra: Construct Complex Table with 'kable' and Pipe Syntax}, 2024.
\newblock URL \url{https://CRAN.R-project.org/package=kableExtra}.
\newblock R package version 1.4.0.

\end{thebibliography}

\address{%
Jayani P. Gamage\\
Monash University\\%
Department of Econometrics and Business Statistics, VIC 3800 Australia\\
\url{https://jayanilakshika.netlify.app/}\\%
\textit{ORCiD: \href{https://orcid.org/0000-0002-6265-6481}{0000-0002-6265-6481}}\\%
\href{mailto:jayani.piyadigamage@monash.edu}{\nolinkurl{jayani.piyadigamage@monash.edu}}%
}

\address{%
Dianne Cook\\
Monash University\\%
Department of Econometrics and Business Statistics, VIC 3800 Australia\\
\url{http://www.dicook.org/}\\%
\textit{ORCiD: \href{https://orcid.org/0000-0002-3813-7155}{0000-0002-3813-7155}}\\%
\href{mailto:dicook@monash.edu}{\nolinkurl{dicook@monash.edu}}%
}

\address{%
Paul Harrison\\
Monash University\\%
MGBP, BDInstitute, VIC 3800 Australia\\
\textit{ORCiD: \href{https://orcid.org/0000-0002-3980-268X}{0000-0002-3980-268X}}\\%
\href{mailto:paul.harrison@monash.edu}{\nolinkurl{paul.harrison@monash.edu}}%
}

\address{%
Michael Lydeamore\\
Monash University\\%
Department of Econometrics and Business Statistics, VIC 3800 Australia\\
\textit{ORCiD: \href{https://orcid.org/0000-0001-6515-827X}{0000-0001-6515-827X}}\\%
\href{mailto:michael.lydeamore@monash.edu}{\nolinkurl{michael.lydeamore@monash.edu}}%
}

\address{%
Thiyanga S. Talagala\\
University of Sri Jayewardenepura\\%
Department of Statistics, Gangodawila, Nugegoda 10100 Sri Lanka\\
\url{https://thiyanga.netlify.app/}\\%
\textit{ORCiD: \href{https://orcid.org/0000-0002-0656-9789}{0000-0002-0656-9789}}\\%
\href{mailto:ttalagala@sjp.ac.lk}{\nolinkurl{ttalagala@sjp.ac.lk}}%
}

\end{article}

\end{document}